\documentclass[12pt,a4paper]{article}
\usepackage{amsmath}
\usepackage{graphicx}
\usepackage{times}
\begin{document}
\begin{titlepage}
\title{Anisotropic flow in pp--collisions at the LHC }
\author{S.M. Troshin, N.E. Tyurin\\[1ex]
\small  \it Institute for High Energy Physics, NRC ``Kurchatov Institute''\\
\small  \it Protvino, 142281, Russian Federation}
\normalsize
\date{}
\maketitle

\begin{abstract}
 We discuss   collective effects in  $pp$--collisions at the LHC energies and derive an upper bound  for the 
 anisotropic flow coefficients $v_n$.  A possibility of its verification via comparison with the measurements of $v_2$ is considered.
 We use an  assumption on the relation  of the two--particle correlations  with
the rotation of the transient state of matter. 
\end{abstract}
\end{titlepage}
\setcounter{page}{2}

\section*{Introduction}
It has appeared  that the interaction of the protons exhibits similarity with interaction of  nuclei, namely, the presence of certain collective effects was  revealed  in  both cases. This point needs to be specified.
The ridge-like  structure has been  observed in correlation function of the two secondary particles  at RHIC in the peripheral collisions 
of nuclei  (cf. for details  \cite{rhicr} and references therein). 
It appeared  that the two-particle correlations of the produced particles have
 a narrow  distribution over $\Delta\phi$  (the relative azimuthal angle of the transverse momenta of the two particles) but wide distribution over $\Delta\eta$  (the pseudorapidity difference of the two detected particles). This phenomenon  called a ridge effect  is usually associated
 with collective properties of a medium produced under interaction of the nuclei. 
 
The similar effect has also been revealed by
the CMS Collaboration \cite{ridgecms} in $pp$--collisions at $\sqrt{s}=7$ TeV in the events with high multiplicities 
and it becomes evident that the form of
 correlations  in pp-collisions resembles the form observed in AA-collisions. Later on, the ridge has 
been found  in nuclear interactions at the LHC, in  PbPb-collisions by  ALICE, ATLAS and CMS \cite{al,at,cm}.
We would like to emphasize here
 that  inelastic peripheral  collisions\footnote{It implies a presence of the  nonvanishing orbital angular momentum in the initial collision events leading to particle production.}
 are present  in  AA--collisions since the ions are extended objects.  Inelastic peripheral collisions in $pp$ case are enhanced  due to a  reflective scattering mode \cite{intja} gradually turning on at the LHC. This enhancement is further amplified by the rescattering in pPb-collisions where the ridge of a significantly higher magnitude that was observed 
 at $\sqrt{s_{NN}}=5.02$ TeV \cite{cmpb}. 

Of course, one should expect a significant quantitative difference between $pp$-, $pA$ and $AA$-collisions and the signal of the ridge and other collective effects should be even much more conspicuous  in $AA$-interactions due to a large size of both colliding objects.  

The experimental results of RHIC and LHC on  ridge in the two-particle correlation functions have 
demonstrated that the emergent hadronic matter  is strongly correlated
and reveals high degree of coherence.  The similarities between the proton and nuclear
collisions have been discussed for a long time ( cf. e.g. \cite{weiner}, \cite{snigirev}). 

It would be helpful to specify  what the term of the reflective scattering mode means.
In fact, the unitarity relation written in the impact parameter representation
implies existence of the two scattering modes, which could be designated as the absorptive and the reflective ones. 
An attractive feature of the impact parameter representation is a diagonalization of the unitarity equation for the elastic scattering amplitude $f(s,b)$, i.e.  
\begin{equation}\label{unit}
 \mbox{Im} f(s,b)=|f(s,b)|^{2}+h_{inel}(s,b)
\end{equation}
at high energies with ${\cal O}(1/s)$ precision \cite{gold}, where $b$ is an impact parameter of the colliding hadrons. The term $|f(s,b)|^{2}$ is the elastic channel contribution, while the inelastic overlap function $h_{inel}(s,b)$ 
covers the contributions from the all intermediate inelastic channels.
An elastic scattering $ S$-matrix element  is related to the amplitude $f(s,b)$ by the relation $S(s,b)=1+2if(s,b)$ and
can be presented in the form
\[S(s,b)=\kappa(s,b)\exp[2i\delta(s,b)]\]
with the two real functions $\kappa(s,b)$ and $\delta(s,b)$. The function $\kappa$ ($0\leq \kappa \leq 1$) is a transmission factor, its value $\kappa=0$ corresponds to  complete absorption. 
At high  energies the real part of the scattering amplitude is small and can  be neglected,  allowing the substitution $f\to if$. 

The choice of elastic scattering mode, namely, absorptive  or reflective one, is governed by the phase $\delta(s,b)$. The common assumption is that  $ S(s,b)\to 0$  at the fixed impact parameter $b$ and $s\to \infty$. It is called a black disk limit, and  in this case the elastic scattering  is completely absorptive. This implies the limitation $f(s,b)  \leq 1/2$.  There is   another option: the function 
$S(s,b)\to -1$ at fixed $b$  and $s\to \infty$, i.e.  $\kappa \to 1$ and $\delta = \pi/2$. This limiting case is interpreted as  a pure reflective scattering \cite{intja}. The principal point here is that  $1/2  < f(s,b) \leq 1$, as allowed by unitarity \cite{phl}.

It is known that to probe experimentally the collective effects, one can use the anisotropic flows coefficients $v_n$  \cite{ollit}.  At the beginning
we obtain an upper bound for the anisotropic flows which is based on the rational form of unitarization and discuss possible
 experimental measurements of the anisotropic flows in the second part of the note. 

\section{Upper bound for the anisotropic flows in $pp$-collisions}
 An appearance of the reflective 
scattering mode at the LHC energies is a key  point for the derivation of the upper bounds for the anisotropic flows coefficients.   
The  solution of the equation $S(s,b)=0$ separates the region where scattering  is a pure absorptive one and the region where reflective scattering is present. 
It corresponds to the maximum value of $h_{inel}(s,b)=1/4$, and the derivative of $h_{inel}(s,b)$ has the form
\[
 \frac{\partial h_{inel}(s,b)}{\partial b}=S(s,b)\frac{\partial f(s,b)}{\partial b},
\]
i.e. it equals to zero at $S(s,b)=0$.  The sign of the derivative of the inelastic overlap function is opposite to the sign of $\partial f(s,b)/\partial b$  when $S(s,b)<0$.
Thus,  the central impact--parameter profile of the function $f(s,b)$ transforms into a peripheral one for the inelastic overlap function $h_{inel}(s,b)$. 
The same can be shown in a different way, namely, the function $h_{inel}(s,b)$ can be expressed as a product, i.e
\[
h_{inel}(s,b)=f(s,b)[1-f(s,b)].
\]
If $f(s,b)>1/2$ at high energies and small impact parameters, then the function $h_{inel}(s,b)$ will have a maximum value of 1/4 at  non-zero impact parameter value.
The inelastic overlap function  in the reflective scattering mode ($S(s,b)<0$) is affected by  the
self-damping of the inelastic channels contribution  \cite{bblan}.
Thus, due to the reflective scattering  the inelastic overlap function $h_{inel}(s,b)$,
\[
h_{inel}(s,b)\equiv\frac{1}{4\pi}\frac{d\sigma_{inel}}{db^2},
\]
has a peripheral  dependence on the impact parameter with a peak at $b=r(s)$. Numerically, it happens in the region starting with energy  $\sqrt{s}\simeq 2$ TeV  \cite{intja}.

A peripheral dependence of $h_{inel}(s,b)$ 
asymptotically  leads to
the approximate relation
for  any observable $A$, which describe a multiparticle production process.  The relation
(valid in the limit $s\to\infty$) has the form  \cite{centr}:
\begin{equation}\label{mst}
A (s,\xi)\simeq A(s,b,\xi)|_{b=r(s)}
 \end{equation}
 since $A(s,\xi)$ ($\xi$ is a variable or a set of  variables), 
can be obtained from the corresponding impact-parameter dependent function $A(s,b,\xi)$
by integrating it with the weight function  ${d\sigma_{inel}}/{db^2}$. In Eq. (\ref{mst}) the function $r(s)$ is determined by the relation $S(s,b=r(s))=0$. Thus, at $b=r(s)$, by the definition of the function  $r(s)$,  the complete absorption of 
the initial elastic channel takes place. 

Eq. (\ref{mst}) attributes the main role to the collision geometry and is applicable for the observables associated with the particle production processes at  $s\to\infty$, where  the reflective scattering being a dominating mode. The energy evolution from the central to the peripheral profile of the inelastic overlap function illustrates this point (Fig.1).
 \begin{figure}[h]
\begin{center}
\resizebox{8cm}{!}{\includegraphics*{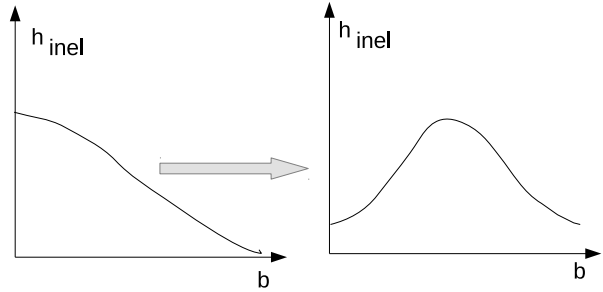}}
\end{center}
\caption[ch1]{\small Evolution of the inelastic overlap function from a central to a  peripheral profile  with increasing center-of-mass energy.}
\end{figure}
 In practice, Eq. (\ref{mst}) can be used with certain precaution for the mean multiplicity, average transverse momentum, anisotropic flows coefficients $v_n$ and multiplicity
 distribution $P_n(s)$ where the reflective scattering becomes noticeable, i.e. starting with the LHC energies. In general,  the relative range of the variations of the impact parameter in the multiparticle production processes 
 decreases with energy and the most typical inelastic event at very high energy is the event with a non-zero value of the impact parameter in the region in the vicinity of $b=r(s)$.  The inelastic events at small and large impact parameter values are strongly  suppressed at very high energies. The energies, where suppression of the inelastic production in the head-on collisions is strong, lie beyond the LHC energy range.
 
 After  these preliminaries, we consider a bound for the anisotropic flows. 
 There are several experimental probes  of collective dynamics. A  most widely discussed ones are the anisotropic flows  \cite{ollit} 
 are determined by the equation
\begin{equation}\label{vn}
v_n(p_T)\equiv \langle \cos(n\phi)\rangle_{p_T},
\end{equation}
which is the n-th Fourier moment of the azimuthal angle distribution of the particles
with a  fixed value of $p_T$.
The angle $\phi$ is
the azimuthal angle of the detected particle transverse momentum with respect to the reaction
plane, i.e. the plane spanned by the collision axis $z$ and the impact parameter vector $\mathbf b$. 
Averaging is taken over large number
 of the inelastic events.
By definition, the absolute value of coefficients $v_n(p_T)$ cannot exceed unity. It is a trivial upper bound. We will
show in what follows that this bound can be reduced by factor of 4 due to  account of unitarity for the inclusive
cross--section.

The inclusive cross-section for unpolarised particles
being integrated over impact parameter $\mathbf b $,  does not depend on the
azimuthal angle of the detected
particle  transverse momentum. It can be written with account for $s$--channel unitarity
in the following form
\begin{equation}
\frac{d\sigma}{d\xi}= 8\pi\int_0^\infty
bdb\frac{I(s,b,\xi)}{[1+U(s,b)]^2}\label{unp}.
\end{equation}
 Eq. (\ref{unp}) has been obtained in the $U$-matrix approach to unitarity when
 the elastic scattering $S$--matrix element is written in the rational form
 in the impact
parameter representation:
\begin{equation}
S(s,b)=\frac{1-U(s,b)}{1+U(s,b)}, \label{um}
\end{equation}
where $U(s,b)$ is the generalised reaction matrix element, corresponding to the elastic $2\to 2$ scattering. It is
considered to be an input dynamical quantity similar to the respective
eikonal function. Since we consider the case of the pure imaginary scattering amplitude, the function $U$  should be pure imaginary too, i.e. the replacement
$U\to iU$ has to be performed.  The function $I(s,b,\xi)$ is an analog of the function $U(s,b)$ for  reactions with particle production (cf. \cite{multrev} and references therein) and $\xi$ is the set of kinematical variables ascribing the final detected particle.

The absolute value and direction of the vector $\mathbf b $ is the main issue under the determination of the anisotropic flows in $pp$--collisions.
This vector can be controlled experimentally in the heavy ion collisions, but the situation, in general, is not explicitly clear
in proton collisions.  However, when we are approaching  the high energy
limit the magnitude $b=|\mathbf b|$ tends to be fixed, cf. Eq. (\ref{mst}), at $b=r(s)$. 
This fact and unitarization allows one to reduce upper bound for $v_n$  by the factor of 4. 
 It should be noted that the impact parameter
$ \mathbf {b}$ is the  variable conjugated to the transferred momentum
$ \mathbf {q}\equiv \mathbf {p}'_a-\mathbf {p}_a$ between two incident channels
 which describes production processes
of the same final multiparticle state.
 
In the case when the impact parameter vector $ \mathbf {b}$ and transverse momentum ${\mathbf  p}_T$
of the detected particle are known
the function $I$ in Eq. (\ref{unp}) does
  depend on the azimuthal angle $\phi$ between two
 vectors $ \mathbf b$ and ${\mathbf  p}_T$.
The dependence on the azimuthal angle $\phi$ can be written in explicit form through the Fourier
series expansion
\begin{equation}\label{fr}
I(s,\mathbf b, y, {\mathbf  p}_T)=\frac{1}{2\pi}I_0(s,b,y,p_T)[1+
\sum_{n=1}^\infty 2\bar v_n(s,b,y,p_T)\cos n\phi].
\end{equation}
The function $I_0(s,b,\xi)$ satisfies  to the
following sum rule
\begin{equation}\label{sumrule}
\int I_0(s,b,y,p_T) p_T d p_T dy=\langle n \rangle (s,b) U(s,b),
\end{equation}
where $\langle n \rangle (s,b)$ is the mean multiplicity produced in collision with the impact parameter $b$.
Therefore, the bare anisotropic  flow coefficients $\bar v_n(s,b,y,p_T)$ are related to the
measured those $v_n$  as follows
\[
v_n(s,b,y,p_T)=w(s,b)\bar v_n(s,b,y,p_T).
\]
where the function $w(s,b)$ is
\[
w(s,b)\equiv [1+U(s,b)]^{-2}=[1+S(s,b)]^{2} /4.
\]
The variable $y$ denotes rapidity, i.e. $y=\sinh^{-1}(p/m)$,
where $p$ is a longitudinal momentum of the particle.

According to  Eq. (\ref{mst}), we should fix the impact parameter value at $b=r(s)$ in the high energy limit.
Since\footnote{Note that $S(s,b)|_{b=r(s)}=0$.} $w(s,b)|_{b=r(s)}=1/4$ and
\[
                   \bar v_n(s,b,y,p_T)\leq 1                                                         
                                                                           \]
 we obtain
 \begin{equation}\label{fr4} 
v_n(p_T)\leq 1/4,
\end{equation}
valid in the high energy limit. This relation provides information on the possible magnitude
of the anisotropic flow coefficients in proton collisions. In the case of $pp$-scattering the respective experimental data
are not yet available, but in the nuclear collisions the  values of  $v_2$, for example, at various energies are
 in the region $0.1-0.2$ at $p_ T \sim 1.5$ GeV/c, i.e. not far below than 0.25. The size of $v_n$ is evidently an energy-dependent 
 in the case of $pp$-collisions in the approach and one could expect its increase with energy since orbital angular momentum increases  with energy too. Therefore, the bound restricting this increase of  $v_n$ can be rather useful.

The natural question is on the possibility to check the above bound experimentally, 
i.e. how to measure $v_n$ in the experiments with  high energy proton beams. We will address this question 
considering for an example the case of the elliptic flow coefficient $v_2$. It is  closely related to the 
reaction plane fixing in the particular inelastic event and will be discussed in the next section.

\section{The ridge and possibility to measure the anisotropic flow coefficients at the LHC with protons}
Elliptic flow coefficient $v_2$ can be expressed  in a covariant form in
 terms of the impact parameter  and transverse momentum vectors
 as follows
 \begin{equation}\label{v2a}
v_2(p_T)=\langle \frac{(\hat{\mathbf  b}\cdot {\mathbf  p}_T)^2}{p_T^2}\rangle-
\langle\frac{(\hat{\mathbf  b}\times {\mathbf  p}_T)^2}{p_T^2}\rangle ,
\end{equation}
where $\hat{\mathbf  b}\equiv \mathbf  b /b$. 

It is useful to recall what is known about this observable from results of nuclear collision experiments.
The differential elliptic flow coefficient $v_2(p_T)$ increases with $p_T$
at small values of transverse
momenta, then it becomes flatter in the region of the intermediate transverse
momenta and decreases at large $p_T$. The integrated elliptic flow coefficient $v_2$  at high energies
is positive and increases with $\sqrt{s_{NN}}$.

 The particular production mechanism leading to appearance of the ridge  \cite{intje} will be  used in what follows. 
 This mechanism is based on the
geometry  of the overlap region in proton collisions and dynamical properties of
the produced transient state\footnote{Of course, this is not a unique mechanism leading to the appearance of the ridge, 
cf.  \cite{rev}}. 
It assumes   deconfinement at the initial stage of interaction. The  
geometrical picture of hadron collision  at non-zero impact parameters
implies \cite{multrev}  that the generated massive
virtual  quarks in the overlap region  could obtain a large initial orbital angular momentum
at high energies. Due to the presence of the strong interaction
between the  quarks this orbital angular momentum would lead to a coherent rotation
of the quark system located in the overlap region. The model estimates of the magnitude of this rotation are based on 
the following assumptions \cite{intje}. The rotation plane coincides with the reaction plane spanned over 
vectors $\mathbf b $ and  the initial particle momentum. It
is similar to the  rotation of the liquid
where strong correlations between particles momenta exist and make their momenta to be coplanar. 
Therefore, the non-zero orbital angular momentum would be realised  as a coherent rotation
of the transient state  as a whole. This state is supposed to be a quark-pion liquid. The arguments in favour
of this claim can be found in \cite{intje}. Finally, the hadronization dynamics
forms a colorless multiparticle  final state.  

The essential point needed  for the existence of  the
 matter rotation is a non-zero impact parameter in the collision.  Of course, this rotation as well as the ridge effect coexist with
 other particle production mechanisms and rotation contributes to the $x$-component\footnote{This component is directed along the vector $\mathbf b $.} of the transverse momentum, i.e. $p_x=p_0+ \Delta p_x$ and does not
 contribute to the $y$-component, $p_y=p_0$.

 It can also be supposed that the constituent quark number scaling \cite{voloshin} observed in heavy-ion collisions at intermediate
 values of $p_T$ and reflecting quark coalescence under the hadronization is to be valid in the proton collisions too, i.e. the following
 relation takes place
  \begin{equation}\label{v2q}
   v_2(p_T)\simeq n_Qv_2^Q(p_T/n_Q),
  \end{equation}\
  $n_Q$ stands for the number of valence quarks in the final hadron and $v_2^Q$ is the elliptic flow coefficient of the quark Q.
The above relation can provide indication on the possible values of the constituent quark elliptic flow in the region
of the intermediate $p_T$ values. Assuming validity of Eq.  (\ref{v2q}), the upper bound 1/12 can be obtained for $v_2^Q$ from the
upper bound for $v_2$.
The  scaling given by the  Eq.  (\ref{v2q}) can be extended into the region of small values of $p_T$ if $v_2$ is plotted
versus transverse kinetic energy $KE_T=m_T-m$ (cf. \cite{he}) and is closely related to the mechanism of hadron formation
in the final state.

As it was noted earlier, the peripheral nature  of the inelastic  $pp$--interactions is controlled dynamically
by the reflective scattering mode. This mode gradually turns on at the LHC energies \cite{intja,degr,alkin}.  Its appearance depends
on the collision energy.
But, the form of $h_{inel}(s,b)$ is
only slightly different from a central one at the center-of-mass energy $\sqrt{s}=7$ TeV. 
The value of $r(s)$  is about 0.2 fm at this energy. Thus, since the highest number of particles are  produced at the small impact parameters\footnote{This is a standard assumption of the geometrical approach.}, the ridge is observed in the events which have high multiplicity at $\sqrt{s}=7$ TeV. With increase of the center-of-mass energy, the maximum of 
$h_{inel}(s,b)$ will be shifted
to the higher values of $b$ and form of $h_{inel}(s,b)$ would become more peripheral. This would lead to transition  
of the ridge to  the events with average multiplicities.Ê This prediction can be tested at the LHC energy $\sqrt{s}=13$ TeV. 

It should also be noted, that  ridge effect can not be observed at the energies $\sqrt{s}\leq 2$ since there is no reflective scattering  in the $pp$-collisions at such energies. The same reason leads to the vanishing coefficients of the anisotropic flow $v_n\simeq 0$ in the 
the $pp$- (and $pA$-)collisions at those energies. This is a distinctive feature of this mechanism. It can be tested  at RHIC.

The idea of rotating transient state can serve as a possible qualitative interpretation of the
ridge and double-ridge structure observed  by CMS, ALICE and ATLAS. 
The narrowness of the two-particle correlation distribution over the azimuthal angle is an important
feature of the mechanism. However, it does not mean that the other options are not able to reproduce
this dependence.  It would be useful to have a quantitative method to analyze the magnitude of the rotation of the system.

It was said that the rotation plane  
 coincides with the reaction plane. Thus, one can determine the normal to this plane being the  unit vector 
\begin{equation}\label{normal}
 \hat{\mathbf  n}=\frac{{\mathbf  p}_1\times {\mathbf  p}_2}{|{\mathbf  p}_1\times {\mathbf  p}_2|}
\end{equation}
Since vectors $\hat{\mathbf  n}$ and $\hat{\mathbf  b}$ are orthogonal, one can rewrite the 
elliptic flow in the form
\begin{equation}\label{v2b}
v_2(p_T)=-\langle \frac{(\hat{\mathbf  n}\cdot {\mathbf  p}_T)^2}{p_T^2}\rangle+
\langle\frac{(\hat{\mathbf  n}\times {\mathbf  p}_T)^2}{p_T^2}\rangle ,
\end{equation} The only experimentally measurable  quantities enter Eq. (\ref{v2b}). The vectors ${\mathbf  p}_1$ and  ${\mathbf  p}_2$ are the momenta of particles 
whose distribution leads to the ridge effect.  Therefore, tagging the  events corresponding to the ridge effect should be performed  during  the analysis. 
Thus, the three-particles' correlations obtained during second stage of the analysis could be sensitive to and provide an information on the elliptic flow coefficient $v_2$.
The above consideration has a  shortcoming since
it is based on the particular model explanation of the ridge effect in $pp$--collisions. However, it allows to estimate roughly  
the values of $v_2(p_T)$, i.e.
together with the upper bound for the flow coefficients one can get hints on the magnitude of the collective effects at the LHC.  The above bound is the main issue discussed in this note.

It should be noted that since the rotating matter involves 
electrically charged constituent quarks, one can expect that direct photons are to be produced by the relativistic synchrotron radiation mechanism and will have similar values of the anisotropic flows coefficients as the respective ones of the secondary
hadrons. The  rotation of the charged quasiparticles could give  a straightforward way to the large elliptic flow of the direct photons at the LHC energies. 

The use of the relativistic synchrotron radiation to explain direct photon elliptic flow at RHIC was suggested in \cite{golov}. As a dynamical reason for such radiation it was proposed to consider confinement phenomena.

\section*{Conclusion}
The main point we would like to stress here is that
the obtained  upper bound for the anisotropic flows can serve as an indicator of the possible magnitude of these observables in $pp$-collisions.
We would also like to note that the ridge effect is expected to become more conspicuous. This is due to the expected 
 growing peripherality of the inelastic collisions  with the center-of-mass energy.  According to the above consideration, this effect should be observed in the events with multiplicities which are more close the the average values as the center-of-mass energy increases.  One could expect that this transition to lower multiplicities
 can be exhibited  already at $\sqrt{s}=13$ TeV where  higher luminosities would be reached in $pp$-collisions. This is a distinctive
 feature of the ridge effect explanation. 
 The upgrade of the LHC would provide additional possibilities for studying the collective effects in $pp$--collisions and the proposed mechanism. 
\section*{Acknowledgement}
We are grateful to A.M. Snigirev for  bringing to our attention the phenomena of the large direct photon elliptic flow.

\end{document}